\documentclass[authorversion,sigconf]{acmart}
\pdfoutput=1
\usepackage{boxedminipage}
\usepackage{xspace}
\usepackage{tabularx}
\usepackage{multirow}
\usepackage{arydshln} 

\usepackage{hyperref} 
\hypersetup{
	pdftitle={ML+FV=<3? A Survey on the Application of Machine Learning to Formal Verification},
	pdfauthor={Moussa Amrani and Levi Lucio and Adrien Bibal}
}

\newcommand{\FV}{\textsc{Fv}\xspace}
\newcommand{\ML}{\textsc{Ml}\xspace}
\newcommand{\TP}{\textsc{Tp}\xspace}
\newcommand{\MC}{\textsc{Mc}\xspace}
\newcommand{\SA}{\textsc{Sa}\xspace}
\newcommand{\SAT}{\textsc{Sat}\xspace}
\newcommand{\etal}{\emph{et al.}\xspace}
\newcommand{\ATP}{\textsc{Atp}\xspace} 
\newcommand{\ATPs}{\textsc{Atp}s\xspace}
\newcommand{\ITP}{\textsc{Itp}\xspace} 
\newcommand{\ITPs}{\textsc{Itp}s\xspace} 
\newcommand{\FOL}{\textsc{Fol}\xspace}
\newcommand{\HOL}{\textsc{Hol}\xspace}
\newcommand{\heart}{\ensuremath\heartsuit}
\newcommand{\RQ}{\textsc{Rq}\xspace}

\usepackage{balance}
\usepackage{inputenc}
\inputencoding{utf8}

\acmDOI{}

\renewcommand\footnotetextcopyrightpermission[1]{} 
\setcopyright{none}

\begin{document}

\title[ML + FV = \texorpdfstring{$\heart$}?]{ML + FV = \texorpdfstring{$\heartsuit$}{}?\\  A Survey on the Application of Machine Learning to Formal Verification} 


\author{Moussa Amrani}
\affiliation{%
  \institution{University of Namur, Faculty of Computer Science, PReCiSE / NaDI}
  \streetaddress{Rue Grangagnage, 21}
  \city{Namur, Belgium}
  \postcode{5000} 
}
\email{Moussa.Amrani@unamur.be}

\author{Levi L\'ucio} 
\affiliation{%
  \institution{fortiss GmbH}
  \streetaddress{Guerickestra\ss e 25}
  \city{M\"unchen, Germany}
  \postcode{80805} 
}
\email{lucio@fortiss.org}

\author{Adrien Bibal}
\affiliation{%
  \institution{University of Namur, Faculty of Computer Science, PReCiSE / NaDI}
  \streetaddress{Rue Grangagnage, 21}
  \city{Namur, Belgium}
  \postcode{5000}
}
\email{Adrien.Bibal@unamur.be}

\renewcommand{\shortauthors}{M. Amrani \emph{et al.}} 
\newcommand\nocell[1]{\multicolumn{#1}{c|}{}}

\begin{abstract}
Formal Verification (\FV) and Machine Learning (\ML) can seem incompatible due
to their opposite mathematical foundations and their use in real-life problems:
\FV mostly relies on discrete mathematics and aims at ensuring correctness; \ML
often relies on probabilistic models and consists of learning patterns from
training data. In this paper, we postulate that they are complementary in
practice, and explore how \ML helps \FV in its classical approaches: static
analysis, model-checking, theorem-proving, and \SAT solving.
We draw a landscape of the current practice and catalog some of the most
prominent uses of \ML inside \FV tools, thus offering a new perspective on \FV
techniques that can help researchers and practitioners to better locate the
possible synergies. We discuss lessons learned from our work, point to
possible improvements and offer visions for the future of the domain in the
light of the science of software and systems modeling.
%
\end{abstract}

%
%

\ccsdesc[500]{General and reference~Validation}
\ccsdesc[500]{General and reference~Verification}
\ccsdesc[500]{Computing methodologies~Machine learning}

\keywords{Formal Verification, Machine Learning}

\maketitle

\section{Introduction}
\label{sec:Introduction}

Formal Verification (\FV) aims at guaranteeing correctness properties of software and 
hardware systems. In that sense, a system is safe with respect to the checked properties. 
Machine Learning (\ML) aims at learning patterns from training data for various
purposes; the derived model generalizes from the data it was trained on.
Both \FV and \ML are grounded on solid mathematical foundations: the former uses
mostly discrete mathematics, fixpoints and abstractions to specify
(concrete/abstract) semantics, properties of interest and the checking process
itself; the latter uses in general continuous mathematics and/or probability
theory to infer models. While they seem at first sight not suitable for each
other, their relative and apparent opposition provides, just like in real life,
the spark for a strong love story.  
This paper focuses on one part of the love
story: what \ML brings to \FV to make it flourish, become more efficient and
accurate, and face real-life challenges? While we are
aware that the topic is broad and the ways in which \ML can help
\FV are necessarily disparate, \FV newcomers and practitioners have
currently no pointers to introduce them to the topic. 

This paper is an attempt to provide a comprehensive survey of the various ways \ML 
contributes to enhance \FV tools' efficiency. To achieve this goal, we propose to
catalog the challenges \FV faces that may be handled through \ML techniques, 
called \emph{themes}, and characterize each theme with a corresponding \ML 
\emph{task}, i.e. \ML problem categories (such as classification, regression, 
clustering, etc.), pointing for each theme to the relevant literature. To the best 
of our knowledge, no contributions in the literature currently exist that spans 
all the spectrum of the main \FV approaches (namely, Model-Checking, Theorem-Proving,
Static Analysis, and to a certain extend, \SAT-solving). By covering various approaches,
we aim at extracting valuable, transversal lessons about general trends of \ML
usage within \FV, as well as provide a high-level snapshot of the current practice in
each \FV approach. 

The main contributions of this paper are the following:
\begin{itemize}
	\item We provide a catalog of themes for each \FV approach, presented in a 
	systematic way: each theme details the corresponding \ML task, and provides
	a commented list of relevant contributions. An overview is available in 
	Table~\ref{tab:summary}.
	
	\item We analyze the literature to extract general observations on the use of 
	\ML inside \FV tools, and to identify some trends and lessons, with an insight
	on what the future may be.
	
	\item We build a comprehensive and searchable repository of contributions found
	in the literature that can help the \FV community build a multi-level 
	understanding of \ML usage in \FV tools. The repository is sorted according to 
	various criteria (publication date, themes, and \ML tasks).
\end{itemize}

This paper is organized as follows. 
Section~\ref{sec:Protocol} describes the paper selection protocol, formulates 
research questions and discusses threats to validity.
Section~\ref{sec:Contributions} analyses in detail the contributions we retrieved.
Section~\ref{sec:Discussion} discusses how our findings answer the
research questions before concluding in Section~\ref{sec:Conclusion}.
 
\section{Background}
\label{sec:Background}

This section provides a high-level introduction to both Formal Verification (\FV) and
Machine Learning (\ML), as the key actors in our survey. For further details, reference
pointers are provided in each section.

\subsection{Formal Verification (FV)}
\label{sec:FV}

In its most classical form, \FV attempts to answer the
following question:
does a \emph{behavioral model}, which reflects the evolution of the various 
variables of a system, satisfy a \emph{specification} of a program, which consists 
in properties of interest characterizing error/undesired states. 

Computing the system's so-called \emph{concrete semantics} explicitly is in the
general case impossible, since it is infinite even for very simple programs.
Rather, \FV proceeds by \emph{abstraction}, or \emph{overapproximation}
\cite{B:Cousot-Cousot:2010}: demonstrating that an abstraction never reaches
forbidden values proves the fact that the actual executions are correct.
However, \emph{false alarms} (or \emph{false positives}) may arise, i.e. errors 
due to an abstraction that is too coarse and that does not correspond to any 
actual system execution. 
Aside from extracting the behavioral model itself, one of the main difficulties 
in \FV is building abstractions that are sufficiently precise to avoid false alarms, 
but sufficiently simple to be automatically computed. This abstraction can take 
on many forms, leading to a variety of \FV approaches.
In this paper we will concentrate on Static Analysis (\SA), Model Checking (\MC)
and Theorem Proving (\TP). We also consider \SAT solving (\SAT): many \FV problems 
can be reduced to the satisfaction of \SAT formul\ae{} \cite{J:Prasad-Biere-Gupta:2005}. 

\subsection{Machine Learning (ML)}
\label{sec:ML}

Humans learn from experience. Car drivers learn by following instructions from
driving school monitors, parents or friends, but also by identifying good
behaviors in other drivers. Players learn chess or basket-ball by studying
``good'' and ``bad'' games, practicing the fundamental moves again and again,
and by identifying best practices that ensure victory. Humans seem also
naturally designed to extract patterns and features in what surrounds them.
For instance, medical doctors provide diagnostics based on many anatomic and
physiological variables such as body temperature or blood pressure -- formally
called \emph{features} in \ML -- available in
patient data, trying to minimize death risk. \ML can be seen as a systematic way
of solving a problem by optimizing some objective function using training data
\cite{mitchell:1997}.

In this paper, we only consider three task categories in \ML: \emph{supervised} and 
\emph{unsupervised learning}, as well as \emph{reinforcement learning}. Other 
categories (like \emph{semi-supervised learning}), as well as many other \ML tasks 
in supervised/unsupervised learning exist, but the ones presented here cover all 
the themes we encountered while analyzing contributions of \ML in \FV approaches. 

In \emph{supervised learning}, a learning algorithm (\emph{learner}) builds a 
predictive model during a training phase, based on features 
found in the training data, by optimizing a defined objective function.
The learned model is then used in a subsequent phase as a predictor for new, previously unseen
data. Tasks can be further classified into problems according to the nature of 
the predicted variable (also called \emph{target}): categorical or continuous. 
For instance, in the medical domain, determining
whether an \textsc{Mri} evidences a cancer is a \emph{classification} task,
because the answer is categorical (a boolean yes/no answer, but more
elaborate classes may be possible); whereas determining which quantity of
insulin should be injected into a patient's blood stream is a \emph{regression}
task, since the predicted variable is continuous.

In supervised learning, the target is known \emph{a priori} and the
learner minimizes some type of distance between the target and its prediction.
On the contrary, in \emph{unsupervised learning}, no target is given \emph{a priori}: 
the \ML algorithm tries to find recurring patterns inside
the data and the final quality judgment is ultimately human. The well-known
\emph{clustering} \ML task consists for instance in grouping elements in the 
dataset, but whether finding three clusters is better than finding five depends on
the domain problem and cannot be answered fully automatically outside the context 
of the algorithm's use. Another task of this kind, the \emph{item set finding} 
task, consists in finding items that may often co-occur together (e.g., buying 
butter and jam may often occur together with buying bread). 

Finally, the third considered category is \emph{reinforcement learning}, which
``is learning what to do — how to map situations to actions — so as to maximize
a numerical reward signal. The learner is not told which actions to take, as in
many forms of machine learning, but instead must discover which actions yield
the most reward by trying them'' \cite{Sutton:1998}.
For instance, an AI agent taking the role of a human player in a computer game
can learn how to complete a level by finding the action patterns leading to a
minimization of penalties.
These penalties may be provided every time the agent dies while trying to
complete a given level of the game.

Once an \ML task is determined, and features of the instances under study are 
identified, an \ML specialist should select the type of \ML \emph{model} that would
approximate the patterns to be found in the data. Examples of such \ML 
\emph{models} are \emph{decision trees}, or the well-known \emph{neural networks} 
\cite{mitchell:1997}. Choosing an appropriate model requires expertise
and fine-tuning (in particular, tuning of the so-called \emph{hyper-parameters}).
As a consequence, failing to obtain meaningful results for a given \ML task may 
mean that the model is inappropriate, or badly parameterized. We
consider that specific considerations on \ML models are beyond the scope of this
paper.
\section{Search Protocol}
\label{sec:Protocol}

\newcolumntype{M}[1]{>{\raggedright}m{#1}}
\hspace{-1cm}
\begin{table*}[ht]%
	\begin{center}
		\begin{tabular}{|M{\columnwidth}|M{\columnwidth}|}
		\hline
		\centering\textbf{FV} & \centering\textbf{ML}\tabularnewline
		\hline
		\begin{scriptsize}
		\textsf{Formal Method} $\lor$ \textsf{Formal Methods} $\lor$ 
\textsf{Formal Analysis} $\lor$ 
\textsf{Formal Verification} $\lor$ 
\textsf{Model Checking} $\lor$ 
\textsf{SAT Solver} $\lor$ \textsf{SMT Solver} $\lor$ 
\textsf{Theorem Proving} $\lor$ 
\textsf{Static Analysis} $\lor$ 
\textsf{Abstract Interpretation} \end{scriptsize} & \begin{scriptsize}
\textsf{machine learning} $\lor$  
\textsf{supervised learning} $\lor$  \textsf{unsupervised learning} $\lor$  \textsf{semi-supervised learning} $\lor$  \textsf{clustering} $\lor$  \textsf{regularization} $\lor$  
\textsf{overfitting} $\lor$  \textsf{underfitting} $\lor$  
\textsf{feature selection} $\lor$  
\textsf{dimensionality reduction} $\lor$  
\textsf{cross-validation} $\lor$  \textsf{backpropagation} $\lor$  \textsf{artificial neural networks} $\lor$  \textsf{deep learning} $\lor$  \textsf{support vector machines} $\lor$  \textsf{kernel methods} $\lor$  \textsf{decision tree} $\lor$  \textsf{decision trees} $\lor$  \textsf{rule learning} $\lor$  \textsf{fuzzy learning} $\lor$  \textsf{meta-parameter tuning} $\lor$  \textsf{hyper-parameter tuning} $\lor$  \textsf{ensemble learning} $\lor$  \textsf{ensemble methods} $\lor$  \textsf{random forest} $\lor$  \textsf{probabilistic learning} $\lor$  \textsf{bayesian induction} $\lor$  \textsf{bayesian probability} $\lor$  \textsf{reinforcement learning} $\lor$  \textsf{regression} $\lor$  \textsf{feature extraction} $\lor$  \textsf{gradient descent} $\lor$  \textsf{cost function} $\lor$  \textsf{data mining} $\lor$  \textsf{data science} $\lor$  \textsf{natural language processing} $\lor$  \textsf{active learning} $\lor$  \textsf{transfer learning} $\lor$  \textsf{matrix factorization} $\lor$  \textsf{manifold learning} $\lor$  \textsf{multidimensional scaling} $\lor$  \textsf{preference learning} $\lor$  \textsf{ranking learning} $\lor$  \textsf{similarity learning} $\lor$  \textsf{distance learning} $\lor$  \textsf{statistical learning} $\lor$  \textsf{density estimation} $\lor$  \textsf{text mining} $\lor$  \textsf{time series analysis} $\lor$  \textsf{predictive model} $\lor$  \textsf{learning bias} $\lor$  \textsf{maximum likelihood} $\lor$  \textsf{k-nearest neighbors} $\lor$  \textsf{k-means} \end{scriptsize} 
\tabularnewline
		\hline
		\end{tabular}
	\end{center}
	\caption{Query String used for Search Engines. We queried the most popular and well-known academic publishers that offer keyword-based search engines (Elsevier ScienceDirect; Springer Link; \textsc{Ieee} XPlore
and \textsc{Acm} Digital Libraries; Semantic Scholar, Scopus, Mendeley and
Google Scholar) with the conjunction of strings appearing in \FV and \ML columns.}
	\label{tab:Queries}
\end{table*}

For realizing this survey, we used a methodology inspired by Kitchenham \cite{TR:Kitchenham:2007}.
Our protocol relies on two observations. First, the authors involved in this work have different, 
complementary backgrounds (the two first authors work on Software Engineering and \FV, 
while the latter is specialized in \ML). Second, no authors had prior knowledge 
of what could exist in the literature: we were quite certain to retrieve only a few 
papers, partly due to the opposition mentioned in Section \ref{sec:Introduction}.
Therefore, we adapted the general methodology of Kitchenham in two ways: we only 
relied on electronic search queries to collect papers (as we were not sure which
academic venues were suitable to such publications, although the study direction
we focus on highly suggested to look into \FV venues); and we conducted a pre-study 
to determine which information is relevant to build our survey. More concretely, 
we followed three steps:
\begin{itemize}
	\item First, we queried several search engines to crawl the largest possible 
	set of relevant publications.
	
	\item Second, we conducted a pre-study on a set of random papers to determine 
	classification categories for analyzing the literature.
	
	\item Third, all authors reviewed the papers and filled a shared document with
	the relevant information extracted from the papers.
\end{itemize}
The rest of this section explains each step in detail, and finishes by formulating 
our Research Questions in Section~\ref{sec:Protocol-RQ}. Table~\ref{tab:summary}
summaries our findings.

\subsection{Search Strategy}
\label{sec:Protocol-Search}

Having no assumption on how to locate relevant papers, we simply opted for a large
list of terms on both sides: we used general-purpose terms for \FV and \ML, strings
corresponding to techniques and algorithms, as well as small variations of those 
terms (e.g., plural and hyphened forms, ``-ing'' forms of verbs, etc.). The search
was conducted between the 10th and the 30th September 2017.
We queried
the main well-known electronic repositories (Elsevier ScienceDirect, SpringerLink,
\textsc{Ieee} XPlore Digital Library, \textsc{Acm} Digital Library, Semantic Scholar,
Scopus and Google Scholar), where we manually processed the result pages and selected 
the relevant publications. 
We discarded some contributions clearly out of scope based on their abstract 
and a quick scan of the content.  
We stopped collecting papers after  
10 pages of results for each search engine, because at that point, most results 
simply correspond to disjunctions of all strings, which becomes highly irrelevant. 
Our search string is formed as a conjunction of the disjunction of the expressions 
in each column of Table \ref{tab:Queries}. Finally, at a later stage, we performed
a lightweight snowballing from the set of papers we collected, in order to retrieve 
papers that may have been missed by our search keywords. This resulted in 264 papers
collected in a shared online repository.

\subsection{Pre-Study \& Paper Filtering}
\label{sec:Protocol-PreStudy}

The next step aimed at discarding clearly irrelevant papers, and performing a 
pre-study to extract analysis categories. Each author selected about 20 papers and
proposed classification criteria that were collegially discussed. We ultimately
retained the elements that constitute a classical \ML pipeline:
\begin{enumerate}
	\item \emph{identifying the theme}, i.e. the \FV problem at hand; 
	\item \emph{identifying the corresponding \ML task}; 
	\item \emph{providing \ML features} to characterize the learning instances; 
	\item \emph{figuring out which \ML model (type)} would perform adequately.
\end{enumerate}
Whether extracting \ML model types (Step 4 of the pipeline) from the papers 
we reviewed has any relevance for readers is debatable: the list we propose is 
informative, since it only reflects the model types authors have selected, but
may in some cases be disputed by \ML experts to be the optimal solution (if
such an optimal solution ever exists).
Nevertheless, we included this information to reflect the literature, such that
readers can grasp what experimentations have been conducted to date for a
particular theme/approach. 

We then performed a first round of reading in order to roughly classify each
paper into \FV approaches, and to discard papers that were clearly out of scope
-- papers that solely focus on one topic (either \FV or \ML), or papers that
leverage dynamic techniques (i.e. that require to actually execute the system). This step resulted
in a categorized repository and a shared spreadsheet for cross-checking papers
that have unclear contributions. When the \FV contribution was not clear (i.e.,
whether it does not fit into an \FV approach), we ensured that a cross-check by
an author with the appropriate background was made; when the \ML
contribution was doubtful (i.e. whether it is really an \ML technique), the 
author with \ML background checked the paper. This resulted in
discarding 96 papers, thus retaining 168 papers for analysis: 53 papers define 
actual themes whereas the rest define auxiliary resources. In particular, we 
list in Table \ref{tab:features} papers that provide
reference contributions regarding the definition of \ML features.

\subsection{Literature Analysis}
\label{sec:Protocol-LiteratureAnalysis}

Once the papers were sorted, the authors with a background in \FV were assigned
two \FV approaches to review: they extracted the theme, identified the corresponding \ML 
task, and retrieved features from each paper. The last author cross-checked the 
most key papers in each approach to allow a comparison from both perspectives,
thus reducing misleading readings about the theme or the task. At later stages when
the list of themes became stable, we made vocabulary used
throughout the approaches homogenous. When possible we factored out the common
features for the approaches (mostly done for \SAT and \TP).

\subsection{Research Questions}
\label{sec:Protocol-RQ}

This survey aims at answering the following Research Questions (\RQ):

\noindent
\textbf{RQ1:} \emph{How is \ML used inside \FV tools?} This \RQ will be answered in
two ways: first, by precisely locating where and how \ML is used inside \FV tools; 
and second, by providing a higher-level overview that spans over all \FV approaches.

\noindent
\textbf{RQ2:} \emph{Is using \ML inside \FV tools beneficial?} This \RQ
is necessary to assess the benefits of \ML in \FV. We answer this \RQ
qualitatively, based on the assessments made by the authors of the papers we
surveyed.

\noindent
\textbf{RQ3}: \emph{What \ML task(s) is (are) used for which purpose in \FV?} 
This \RQ is intended to associate an \ML task to an \FV theme, as
described in Background Section \ref{sec:ML}, and helps bridging both worlds by 
relating activities in \FV tools to a meaningful category for \ML experts.

\noindent
\textbf{RQ4}: \emph{Which model types are used to perform the \ML tasks?}
This \RQ is intended to collect from the reviewed papers the \ML model types the
various authors have used to enhance \FV tools. Although indicative and by no 
means complete or definitive, it provides an interesting panorama of the current 
practice.

\noindent
\textbf{RQ5}: \emph{How are \ML features extracted/selected to guide \ML tasks?}
This \RQ is intended to locate the \ML instances' characteristics as used in the 
literature, and to eventually list the most common ones.

\section{Contributions}
\label{sec:Contributions}

This section catalogs some of the ways \ML complements \FV approaches. 
We aimed at representativity, i.e. we tried to maximally cover the \ML/\FV 
complementarities (called \emph{themes} from now on) to propose an overview of the panorama of techniques and current practices in this field.
To the best of our knowledge, this is the first study that spans over the main \FV approaches
to provide insights on how \ML participates in \FV tools.

We organize this section by \FV approaches in a self-contained way such that
each may be read independently. We start with \SAT-\textsc{Smt} Solving, then
Theorem Proving (\TP), which corresponds to a progression in the expressiveness
of the underlying logics (propositional/boolean, then First-Order and
Higher-Order). After we introduce and Model Checking (\MC), given the
particularity that temporal logic deals with time, and end with Static Analysis
(\SA).

Each approach follows the same outline. First, we briefly recall how the \FV approach
works. Second, we explain the sources of complexity (typically, NP-completeness) 
and which countermeasures (e.g. heuristics) have been historically designed to 
partially overcome them. When necessary, we introduce a brief explanation of the 
main algorithm supporting the approach in order to fix the terminology and to situate 
how each theme finds its place in \FV. Third, we introduce for each theme where
the \FV/\ML complementarity exists, ground it in terms of \ML task(s), and finally provide examples from the literature.
When possible, we indicate the \ML features associated with the theme: when they 
are common, they are factored out into the section's headers; otherwise, they 
appear in each particular theme.

In order to guide the reader, Table \ref{tab:summary} gathers the highlights of 
our findings in a comprehensive way:
for each theme identified within each approach, we gather all the
selected contributions from the literature, the \ML tasks used in that theme and
provide hints on the \ML model types that these contributions used.

\subsection{SAT / SMT Solving (SAT)}
\label{sec:SATSMT}

The \SAT problem is a decision problem: given a boolean propositional formula,
find one (or several) valuation(s) for which the formula evaluates to \textsf{true}. 
When such a valuation exists, the formula is said to be \emph{satisfiable} (and
\emph{unsatisfiable} otherwise). Usually, a formula is processed starting from a
canonical representation such as the Conjunctive Normal Form (\textsc{Cnf}),
where formulas consist of disjunctions of clauses, which are themselves conjunctions
of literals defined as variables or their negation.
Theories may enrich propositional boolean formul\ae{} to represent e.g.
first-order logic, numbers or richer data structures such as arrays or lists, resulting in the Satisfiability
Modulo Theory (\textsc{Smt}) decision problem. Some \textsc{Fv} approaches may
be reduced to a \SAT problems (e.g., \MC \cite{Amla-Du-etAl:2005} and 
\TP \cite{B:Biere-Heule-vanMaaren-Walsh:2009}; for \FV see 
\cite{J:Prasad-Biere-Gupta:2005}), making \SAT research relevant for
\textsc{Fv}.

No algorithm can solve all \SAT instances efficiently, which results in a plethora 
of solving algorithms. Most of the existing algorithms are variations of the
Davis–Put\-nam-Logemann–Loveland (\textsc{Dpll}) algorithm. In practice, 
tools need to carefully choose the adequate variation for a given (set of) instance(s) 
to solve them efficiently, generally by reducing the overall runtime.
In a simplified way, the \textsc{Dpll} algorithm proceeds as follows: first,
chooses a branching literal and assign it a truth value; then, propagates this
assignment to other clauses, resulting in unit clauses, i.e. clauses in which only one
literal remains unassigned, making its assignment obvious; and finally,
propagates those choices appropriately until full assignment, or detection of 
conflict. When facing a conflict the algorithm backtracks to the
previous branching literal to try the opposite assignment. These steps apply recursively
until success or unsatisfiability. The Conflict-Driven Clause Learning
(\textsc{Cdcl}) improves the general \textsc{Dpll} algorithm by analyzing the
cause of the conflicts and backtracking to the appropriate level instead of
simply the previous choice, thus improving the overall runtime. 

Historically, the \SAT community already identified a number of (\SAT) instances
\emph{features} that characterize the hardness of satisfying an instance. 
We detail here the most important ones, and mention in the themes the contributions' 
features that differ from them. 
SATzilla \cite[Fig. 2]{J:Xu-Hutter-Hoos-LeytonBrown:2008} integrates a large number of
these well-recognized features (around 150): instance size metrics; Variable
Incidence / Clause-Variable Incidence Graphs (\textsc{Vig}/\textsc{Civg})
metrics, balance of positive/negative literals in clause and variable occurrences,
binary/ternary and Horn clause fractions, number of unit propagation,
search space size and other local search probing characteristics. 
The contributions not explicitly dealing with feature improvements basically reuse different subsets of these features (see e.g.,
\cite{Horvitz-Ruan-etAl:2001,Wu:2017} among many others). 
For 3-\textsc{Cnf} \SAT instances, the authors of SATzilla \cite{Xu-Hoos-LeytonBrown:2007} 
managed to reduce to five the list of the most prominent features, without significant loss 
of performance.
Ans\'otegui and his co-authors \cite{J:Ansotegui-Bonet-GiraldezCru-Levy:2016} defined new 
interesting features, targeting industrial \SAT
instances: the scale-free structure assumes that the ratio of variable
occurrences and total number of variables follows a power-law distribution; the
community structure measures the modularity of graphs, i.e. how high is a node
connected to its direct neighbors; and the self-similar structure measures the
fractal effect of \textsc{Cig} and \textsc{Cvig}, i.e. how it changes when a
group of nodes is replaced by a single one. They show that relying on those
features is computationally affordable, and predicts the instance satisfiability
in a way that is comparable to taking all features that SATzilla uses. 
\subsubsection{Predicting Runtime}
\label{sec:SATSMT-RuntimePrediction}

Predicting the runtime of an instance (or a subformula) is helpful in many
regards: choosing an appropriate solver depending on the runtime; interrupting a
computation to switch to another algorithm when the current one takes too long;
selecting an appropriate restart strategy when encountering conflicts; selecting
a variable to branch on depending on the runtime that will likely result; etc.
Runtime is a real, continuous value, making this prediction, strictly speaking,
a \emph{regression} \ML task. However, the literature often considers
runtime \emph{classes} (e.g. long/short runtime of a specific
algorithm, i.e. observing whether the runtime crosses a predefined threshold,
considering a time budget for solving an instance set). This results in
practice in a \emph{classification} \ML task.

Horvitz, Ruan \emph{et al.} \cite{Horvitz-Ruan-etAl:2001} estimated the runtime
of the Quasigroup Completion Problem (\textsc{Qcp}, closely related to \SAT) by
defining a set of features that accurately estimates the resolution progress by
reflecting the instance patterns (instance size) and dynamic information about
the solver's state (number of backtracks, search tree depth). Samulowitz and
Memisevic \cite{Samulowitz-Memisevic:2007} targeted Quantified Boolean
Formul\ae{} and proposed various features: the \textsc{Vsids} (Variable State
Independent Decaying Sum) score, the number of conflicts and the fraction of
already solved clauses, the weighted sum between forced literals and
\textsc{Vsids} scores.

\subsubsection{Restarting Computations}
\label{sec:SATSMT-Restart}

When encountering a conflict during resolution, the analysis of the literals
that led to assignment inconsistencies allows to efficiently backtrack to an
appropriate level, avoiding traps that would likely result from a backtrack at
another level. A theoretical instance-specific optimal restart strategy
exists, but requires the knowledge of the instance's runtime distribution
(rarely known and difficult to compute) \cite{Luby-Sinclair-Zuckerman:1993}.
Those strategies are split into two categories: \emph{universal} strategies are defined
independently of the instance; and \emph{dynamic} strategies adapt to the search length,
i.e. the number of already assigned literals. Since no strategy outperforms all
others on all datasets \cite{Huang:2007}, adapting the strategy to a dataset is
often better, and constantly reevaluating the portfolio of restart strategies is desirable to keep improving \SAT solvers performances \cite{Biere-Frohlich:2015}.

For solving as many instances as possible within a given time budget, selecting 
the best restart strategy on an instance basis (according to its features) represents 
a \emph{classification} \ML task: 
from a set of predefined strategies, determine which one would be the best to 
optimally (i.e. by minimizing the expected runtime) solve the instance at hand. 
For very hard instances (typically, industrial/crafted instances representing 
cryptographic or planning problems \cite{J:Ansotegui-Bonet-GiraldezCru-Levy:2016}), 
or for solving several instance sets while minimizing the overall runtime,
switching between strategies during the solving is often more efficient, when some
strategies are known in advance to be the most powerful for the given instances 
(sets). This can be seen as a \emph{reinforcement learning} problem, in which 
restart strategy choices are seen as the possible \emph{actions} to reinforce. Rewards 
are defined differently depending on the particular contributions.

Haim \& Walsh \cite{Haim-Walsh:2009} proposed to select a strategy from a
portfolio of 9 that were proven to perform well on at least one dataset. The training 
is based on a subset of features taken from \cite{J:Xu-Hutter-Hoos-LeytonBrown:2008}; 
the prediction is realized dynamically, while solving the instance. 
Horvitz, Kautz, Ruan and their colleagues adopted a more contextualized
approach, in the context of dependent runs 
\cite{Kautz-Horvitz-Ruan-etAl:2002,Ruan-Horvitz-Kautz:2002}: they used predictors 
(called observators) on some instances of a set to help determine how to perform 
restarts for the other instances of the set. Observators are generally classifiers 
trained on a few instances that predict whether (future) instances would be 
satisfiable or not, or whether their runtime takes a short/long time.

Nejati \emph{et al.} \cite{Nejati-Liang-Ganesh-Gebotys-Czarnecki:2017} targeted 
cryptographic instances, whose runtime is usually much longer than other
instance types, making restart strategies a core component. Solving an instance requires, 
during the solving itself, to change / switch strategies, among the ones that are
known to be the most effective in the literature: uniform, linear, luby and 
geometric \cite{Biere-Frohlich:2015}). 
Reinforcement Learning is performed through the following steps. First, a strategy is 
chosen, and the general algorithm proceeds with the solving until the strategy 
imposes a restart. 
At this point, the strategy is rewarded based on the average Literals Block Distance 
(\textsc{Lbd}) of the learned clauses generated since the strategy was selected.
Finally, this results in choosing/favoring strategies that produce small \textsc{Lbd}s for 
the future solving steps.
Gaglio and Schmidhuber \cite{Gagliolo-Schmidhuber:2007} considered the problem of
using the best restart strategies for a set of instances to minimize the global 
runtime. They choose between two strategies (Luby's universal and uniform). After
one step of solving an instance, they reward the strategy that results in a runtime
that stays close to the runtime of the previous instances. This results in favoring
the strategy that provide a global runtime for the set that is the closest to the 
runtime of most of the instances in the set.

\subsubsection{Selecting the Branching Variable}
\label{sec:SATSMT-Branching}

Choosing the most appropriate branching variable is crucial for improving solvers'
runtimes, because it ultimately minimizes backtracking (which can be seen as a 
step back towards a solution, since it implies unassigning some of the already
selected variables).  
This can be seen as a \emph{reinforcement learning} problem: along a \SAT instance
solving, choose the next variable to branch on, such that the reward attached to
the variable choice, called \emph{score}, maximizes the progress for solving the instance.
Note that this task is \emph{non-stationary} from the \SAT solving viewpoint: after
each choice, a variable cannot be selected anymore unless a conflict occurs.

Liang \emph{et al} 
\cite{Liang-Ganesh-Poupart-Czarnecki:2016a,Liang-Ganesh-Poupart-Czarnecki:2016b} explored 
two different reward computations based on different branching heuristics: in 
\cite{Liang-Ganesh-Poupart-Czarnecki:2016a}, they used a conflict-history-based 
heuristic for variable selection; while in \cite{Liang-Ganesh-Poupart-Czarnecki:2016b},
they used another heuristic called learning rate branching. Both rewarded the 
generation of learned clauses locally, at each step of the solving. 
Later on, Liang \emph{et al.} \cite{Liang-etAl:2017} rewarded selections that 
maximize \emph{global} branching learning rates, i.e. rates for the whole solving.
Fröhlich \emph{et al.} \cite{Frohlich-Biere-Wintersteiger-Hamadi:2015} penalize the 
candidate variables choices that minimizes the number of unstatisfied clauses. 
Lagoudakis and Littman \cite{J:Lagoudakis-Littman:2001} penalize branching rules 
(chosen among seven known as the best working) whose solving time is too long.

\subsubsection{Determining Best Solving Algorithm} 
\label{sec:SATSMT-Best}

The \SAT community identified families of instances that may be better solved
with specific algorithms, enhancing specific criteria (mostly, solving runtime).
In a pure form, this is a \emph{classification} \ML problem: from a set of instances, 
determine which solving algorithm(s) would be the most efficient according to a
given criterion.
However, this could be seen as a \emph{regression} \ML problem, when the goal is to predict an algorithm continuous probability of success. 
Both \ML tasks are achieved offline, i.e. before 
running the solving algorithms. Note that \emph{pre-solving} is mainstream, i.e. 
trying some predetermined, quicker algorithms that may solve some of the instances, 
leaving the portfolio selection to focus on difficult instances. 

SATzilla \cite{J:Xu-Hutter-Hoos-LeytonBrown:2008} is one of the best portfolio
solvers \cite{LeytonBrown-Nudelman-Andrew-McFadden-Shoham:2003IJCAI}: it relies 
on a large variety of specialized \SAT algorithms that are
chosen according to the specificities of the instance at hand, based on 115
features (cf. feature discussion in Section's header). 
It allows to switch to another algorithm (``next
best match'') when the one attributed initially takes too long. AutoFolio
\cite{J:Lindauer-Hoos-Hutter-Schaub:2015} selects algorithms based on features
similar to the ones in SATzilla.

\subsubsection{Configuring \SAT Solvers' Parameters} 
\label{sec:SATSMT-Parameterisation}

Instead of setting default values for the multiple parameters of the various
\SAT algorithms constituting a portfolio, many \SAT solving tools choose to
expose those parameters to the end-users, passing them the burden of
configuration. The end-user faces a highly difficult task: which parameter
settings of the algorithm(s) perform best on a set of instances, minimizing a
cost function (typically in \SAT, runtime). Several approaches already exist
based on heuristics, but the domain recently gained
attention with \ML. This particular domain has its own competitive event: the
Configurable \SAT Solver Challenge. Finding the optimal values of a \SAT
solver's parameters is a {\emph regression} or a {\emph classification} task whether if a continuous or a categorical value is predicted.

Hutter, Hoos and Leyton-Brown \cite{Hutter-Hamadi-Hoos-LeytonBrown:2006} defined 
\textsc{Smac} (Sequential Model-based Algorithm Configuration), a technique and 
tool that generalizes the classical optimization algorithm by using training, 
based mostly on SATzilla's features. 
AutoFolio \cite{J:Lindauer-Hoos-Hutter-Schaub:2015} parametrizes ClasspFolio 2 (the 
default version of SATzilla'11), resulting in a
highly parametrized algorithm framework. 

\subsubsection{Learning Satisfiability}
\label{sec:SATSMT-InstanceSatisfiability}

Tackling the whole \SAT problem through \ML seems difficult, but
has been partially attempted. Overall this corresponds to a \emph{classification} 
problem (although many other \ML tasks are performed in between to serve the 
main goal): determine whether a \SAT instance is satisfiable or not.

A partial prediction on a subformula may guide a solver for other tasks (e.g., 
restarting efficiently, or evaluating the potential of a variable selection, 
among others). For example, Wu \cite{Wu:2017} predicted 3-\textsc{Cnf} instance 
satisfiability with seven features from SATzilla and other classical ones. They 
reuse the partial prediction to determine which value is preferable for a 
branching literal. Although the technique is applied to subformul\ae, nothing 
prevents the technique from being used on a larger scale (although optimizations are 
naturally expected).
One would expect that a \SAT problem becomes harder when the number of clauses 
increases. In fact, the most difficult instances are those whose ratio of clauses
over variables is near the so-called \emph{phase transition} (particularly for 
3-\textsc{Cnf} instances): the number of clauses and variables is at equilibrium,
making instances neither underconstrained (i.e. exhibiting many possible 
solutions), nor overconstrained (i.e. exhibiting many contradictions). 
Devlin and O'Sullivan \cite{Devlin-OSullivan:2008}, and later on Xu, Hoos and Leyton-Brown 
\cite{Xu-Hoos-LeytonBrown:2012} studied several classifiers for predicting the 
(non-)satisfiability of general \SAT as well as 3-\textsc{Cnf} instances, based 
on the classical features used in SATzilla. Xu, Hoos and Leyton-Brown tried to minimize
the number of necessary features to build good classifiers, and managed to reduce
to two features while staying robust comparing to classifiers with more features.

\subsection{Theorem Proving (TP)}
\label{sec:TP}

When the semantics of a software or of a system is expressed as mathematical
theories, verification conditions for those systems can be
formulated as mathematical properties of those theories. In \FV, Theorem Provers
(\TP) are then employed, in a more or less automated fashion, to prove or
disprove such properties.

Mathematical theories are composed of mathematical facts, which are assumed to
be true. \TP is used to infer new facts about the theory, using the inference
rules associated to the logic of choice. In this sense, a mathematical statement
(known as conjecture) becomes a theorem if it logically follows from the theory.
More precisely, \TP operates as follows: 1) it receives as input a set of facts
from a mathematical theory which are assumed to be true and a \emph{conjecture};
2) it performs a number of inferences on those facts using the set of rules that
describe the semantics of the logic being used; and 3) outputs a proof
for the conjecture or a trace thereof, if one exists, in which case the
conjecture becomes a \emph{theorem} and can be added as a new fact to the
theory.

First-Order Logic (\textsc{Fol}), one of the most popular logics in \TP, is
semi-decidable. The bulk of the work of applying \ML techniques to TP has thus
targeted \textsc{Fol}. \ML techniques assist or replace existing expert
knowledge-based heuristics in order to better navigate the border between
decidability and non-decidability and more efficiently lead theorem proofs to
completion. Because they present a high level of automation, provers for
\textsc{Fol} are called \emph{automated theorem provers} (\ATPs).

Higher-order logic (\HOL) adds to \FOL the quantification of predicate
and function symbols. Such expressiveness is convenient to express verification
problems that would otherwise be too difficult or impossible to express in lower-order logics.
However, \HOL is undecidable and presents fewer opportunities for automation
than \FOL, which means parts of the proofs need to be guided by humans. For this
reason, in the context of \HOL, provers are called \emph{interactive theorem
provers} (\ITPs).

Decidability and efficiency issues in \ATPs/\ITPs mean that decisions are
delegated onto humans at many points of the proof.
Such decisions involve for instance: choosing facts (also known as
premises~\cite{KuhlweinLaarhoven:2012}) from the theory relevant to the proof at
hand; picking sets of proof engine parameters (also known as
\emph{heuristics}~\cite{Bridge2010}) such as for instance
sets of inference rules used~\cite{KuhlweinUrban:2015}.
It is in supporting or replacing the human in these decisions that \ML comes to
the aid of theorem proving.

Features used to characterize facts or conjectures about theories are
majoritarily the symbols found in those mathematical
statements~\cite{meng:2009}, for example literals or predicate names. Metrics
such as the number of clauses, literals or subterms, or yet specific metrics
about the translations of logical formulas into normal forms have also been
used~\cite{KuhlweinUrban:2015}. Other authors have attempted to use
types~\cite{Kaliszyk2014}, or meta-information about the theory name and its
presence in various mathematical databases~\cite{Kuhlwein2013}. Recently,
Kaliszyk and his co-authors have proposed features that capture semantic
relationships between mathematical statements~\cite{Kaliszyk:2015}.
 
\subsubsection{Selecting Facts} 

Which subset of facts to take from a large theory in order to complete a
given proof as efficiently as possible (or at all) is one of the most prominent
applications of \ML to \TP~\cite{KuhlweinLaarhoven:2012}. Fact selection is a
\emph{classification} task: either a fact is relevant for the current proof or
not, potentially with a probability reflecting a level of certainty. 

Fuchs~\cite{Fuchs1998} uses data from previous proofs to train a model for
computing the usefulness of the available facts for the next proof step. Alama
\cite{Alama:2011} preanalyzes a large mathematical repository of formalized
theories in order to calculate dependencies between parts of those theories that
can then be used at proof time. Kaliszyk
\etal~\cite{Kaliszyk2014,Blanchette:2016} use classification models to rank
facts in \HOL theories according to their assumed relevance for the proof of the
conjecture. They then reduce the best ranked of those facts to simpler problems
that can be handled by fast \FOL \ATPs to help in parts of the proof of the
original conjecture. Again Kaliszyk, together with his co-authors, provide
in~\cite{KaliszykUV:14} a compelling account of how \textsc{MaLARea} performs
\ML-based fact selection for the equational reasoning \ATP
\textsf{E}~\cite{Schulz:2002} beating the competition in large-theory contests.
Alemi \etal report in~\cite{AlemiCISU:16} the first application of deep learning
to fact selection for large theories, concluding that neural networks do help
in large-scale automated reasoning without requiring hand-crafted features. They
mention nonetheless that hybrid premise-selection solutions where hand-crafted
solutions are used together with their solution may yield even superior results.
Loos and her colleagues confirm this thesis in~\cite{Loos:2017} and conclude
that fact selection mixing neural networks and other methods is particularly
useful for hard theorems that require complex reasoning.

\subsubsection{Configuring Proof Engine Parameters}

Proving a particular conjecture is typically achieved more or less efficiently
(or at all) by providing the prover with a set of parameters. It is well
established in the \TP community that certain parameters configurations 
are better suited for the proof of certain classes of
conjectures~\cite{Bridge2010}. Constructing such parameters automatically for
specific conjectures is assisted by an \ML \emph{regression} task: \ML
helps in predicting proof runtime when heuristics are evaluated on specific
conjectures.

K\"uhlwein and Urban investigate in~\cite{KuhlweinUrban:2015} a method to
automatically tune parameters of \ATPs in order to optimize proof times . Their
method starts from a set of random or predefined heuristics and the \ML
algorithm learns to predict how fast these heuristics perform on classes of
existing problems. The technique then iteratively improves the parameters of
successful heuristics by using the prediction learner.

\subsubsection{Selecting Pre-Defined Proof Engine Parameters}

Some theorem provers select proof engine parameters that were
manually or automatically generated (cf. \emph{Configuring Proof Engine
Parameters} theme). This is a \emph{classification} \ML task: given a
conjecture, provide the heuristic that will most likely produce a proof for it in an efficient manner.

Bridge~\cite{Bridge2010}, a
reference for this theme, evaluates more than fifty features and concludes that only combinations of very few features (up to two)
are required to build classifiers that vastly outperform random proof engine
parameter selection. With his colleagues \cite{Bridge2014} Bridge later confirms
that their system yields the same performance as \textsf{E}'s internal proof
engine parameter selection mechanism, without requiring the introduction of any
human-expert knowledge. Additionally, the system is also able to decline some
proofs in case no proof engine parameters can lead to the completion of the
conjecture's proof in an acceptable amount of time (or at all). The authors report that declining proofs greatly improves performance,
while only moderately reducing the amount of provable theorems.

\subsubsection{Guiding Interactive Proofs}

Due to the undecidability of \HOL, there is no systematic way of finding proofs
for conjectures in such logics. \ITPs such as
\textsc{Coq}~\cite{coq-refman:2009}, \textsc{Isabelle}~\cite{Blanchette:2011} or
\textsc{Mizar}~\cite{Grabowski:2010} are used to assist the mathematician in
proof finding, while \textsc{Proof General}~\cite{Aspinall:2000} provides a
high-level user-friendly interface to those \ITPs. \ITP environments can act as
recommender systems to suggest for example promising facts to be used in the
proof. Because this theme touches many parts of proofs, both
\emph{classification} and \emph{clustering} \ML techniques can be used.
\emph{Clustering} becomes particularly useful here as it can inform the user of
potential next steps through statistical analysis on similar proofs -- it
however cannot lead to automatic decision making such as when supervised
approaches are used.

Urban \cite{Urban:2006} describes a set of proof aids in \textsc{Emacs} for
\textsc{Mizar}, explaining how \ML classification algorithms are used to suggest
facts to a mathematician for the continuation of the proofs. Mercer
\etal \cite{Mercer:2006} propose a system and a user interface for recommending
the next proof step, based on Duncan's work \cite{Duncan:2004} on modeling
proofs with Variable Length Markov models.
Komendantskaya and Heras interface \textsc{Proof General} with back-ends running
clustering algorithms \cite{Komendantskaya:2012} to
gather statistics on data from previous proofs.

\subsubsection{Learning Theorem Proving} 

Rockt\"aschel and Riedel attempted to learn the backward chaining algorithm for
\FOL~\cite{Rocktaschel:2017}. Starting from a set of neural networks that
modularly perform generic \TP-related tasks (such as for example
\emph{unification}), the authors propose an algorithm that is able to assemble
those modules in order to deduce new theorems from a given knowledge
base.
This contribution has the particularity that, due to the fact that modules are
used, the proof is in itself interpretable --  more specifically how those
modules are used during the proof provides a proof trace. As the output of the
neural network is a proof score that describes the confidence in the derived
facts, we technically classify it as a \emph{regression} task.

\subsection{Model-Checking (MC)}
\label{sec:MC}
\enlargethispage{12pt}

Model-Checking (\MC) \cite{Queille-Sifakis:1982,J:Clarke-Emerson-Sistla:1986}
consists of abstracting the concrete system's execution into a finite-state
automaton (that can be extracted automatically from the program, and whose 
execution may be infinite); and the properties of interest are expressed in temporal logics.
The model-checking procedure explores exhaustively the (abstract) state space,
and either validates the properties, or returns a counterexample (that may be a
false alarm).
In practice, exhaustive exploration is difficult: many techniques were crafted
to reduce the state space (symmetry, slicing, partial evaluation, to cite only a few), 
or enhance its exploration (through path exploration heuristics). They nowadays equip 
most tools. We did not find \ML contributions that work at the \MC algorithm level
like for others \FV approaches; rather, we found contributions that help
detecting counterexamples faster, or reducing false positives (using the
well-known \textsc{Cegar} approach).
Note that we did not include contributions for Assume-Guarantee Reasoning
\cite{Cobleigh-Giannakopoulou-Pasareanu:2003,Barringer-Giannakopoulou-Pasareanu:2003,Gheorghiu-Giannakopoulou-Pasareanu:2007,J:Pasareanu-etAl:2008,Chaki-Clarke-Sinha-Thati:2005,Sinha-Clarke:2007,Nam-Alur-Madhusudan:2005,Nam-Alur:2006}
based on the $L^*$ algorithm \cite{J:Angluin:1987,J:Rivest-Schapire:1993}: it is
difficult to conclude without deepening the subject whether $L^*$ is an \ML
algorithm, and which \ML task it corresponds to.
\subsubsection{Finding a Counterexample}
\label{sec:MC-Counterexample}

Model-Checkers should be oriented towards \emph{error detection}
\cite{J:Clarke-Wing-etAl:1996}: they should favor the discovery of errors rather
than focusing on guaranteeing correctness. As a consequence, optimizing
counterexample finding is crucial. A possible
approach is to explicitly \emph{guide} the state space exploration towards paths
that may favor such counterexamples, based on the property of interest at hand.
This may be achieved through \emph{reinforcement learning}: a reward favors
positive paths for invalidating the property; while a punishment discourages
negative paths validating it. Note that the qualificatives \emph{positive}/\emph{negative} 
correspond to the \emph{error detection goal} instead of the traditional \MC goal. 

Araragi and Mo Cho \cite{Araragi-MoCho:2006} targeted the production of
counterexamples for \emph{liveness} properties that represent responses, i.e.
a (premise) event is expected before a (response)
event should eventually occur. The authors kept track of the premise occurrence at the state space level, and rewarded explorations that stayed on paths between premise and response as long as possible. This would lead to cyclic, or very long (or infinite) paths that would invalidate the property.
Behjati, Sirjani and Ahmadabadi \cite{Behjati-Sirjani-Ahmadabadi:2009} studied 
\textsc{Ltl} properties on Büchi automata with on-the-fly \MC: 
the reinforcement learning agent is punished when following non-accepting cycles, 
and rewarded when finding unfair accepting cycles, until a fair one is found, leading to the property's invalidation.

\subsubsection{Refining Abstraction based on Counterexamples (CEGAR)}
\label{sec:MC-CEGAR}

A spurious counterexample (false alarm) happens when the last state of a path in
the concrete system mixes both \emph{deadend} states, i.e. states with no
concrete transition to the failure state; with \emph{bad} states, i.e. states
using (system) variables useful to prove the property, that are not taken into account,
and abstracted together with deadends. To eliminate such a counterexample, some
variables need to be identified and become visible, i.e. separated properly
within the abstraction. This is known as the \emph{separation problem} in
\textsc{Cegar} \MC, which is a \emph{classification} \ML task: given a (sub)set of 
system variables from the failure state, determine whether they 
should be classified as \emph{deadend} or \emph{bad}. This information then 
allows for an abstraction refinement that, even if not optimal, makes it possible to discharge 
the counterexample. Note that for realistic systems, enumerating all the 
variables is impossible: a preselection is generally operated beforehand.

Clarke, Gupta and their colleagues 
\cite{Clarke-Gupta-Kukula-Strichman:2002,J:Clarke-Gupta-Strichman:2004} implemented
this technique for model-checking hardware circuits, training a learner on
samples automatically extracted from the concrete system.

\subsubsection{Extracting Most Common Error Patterns}
\label{sec:MC-Patterns}

Concurrency errors often result from the same error types
\cite{Lu-Park-etAl:2008,Rungta-Mercer:2007}. Finding the recurring paths or
rules leading to such errors is related to the \ML task of \emph{frequent item set finding}.

Pira, Rafe and Nikanjam \cite{J:Pira-Rafe-Nikanjam:2016} characterized frequent
patterns as sequence of rewriting rules in Groove, a graph-based model
transformation tool, using a variation of the APriori algorithm. Those patterns
are discovered on smaller systems with similar architectural design, then used
to guide \MC on larger systems.

\subsection{Static Analysis (SA)}
\label{sec:SA}
\enlargethispage{12pt}

Static Analysis (\SA) designates a large panel of techniques aimed at computing 
any information about a program, generally directly on its Abstract Syntax Tree 
\textsc{Ast}. 
The underlying abstractions rely on predefined approximations (possibly parameterized 
by users' inputs): this results in a
fixed set of properties of interest, e.g. extracting Android apps' permissions 
from \texttt{.apk} files, which may be parameterized to find only device-specific ones.
In most cases, the analysis does not carry in itself its own final usage:
for instance, permissions, in themselves, do not give any information about
an app being a malware. Most of the contributions in \SA leverage \ML to bridge
this gap, by trying to find links, or reccurent patterns, in the retrieved information, 
e.g. malwares are apps that present significant discrepancies between exhibited 
permissions and actual executed code.

\subsubsection{Identifying Actionable Alerts}
\label{sec:SA-Alerts}

\SA tools often issue large amounts of alarms that warn about code style
violations, trivial defects with no impact on functionalities, false positives
and, of course, real bugs. Too many warnings hamper developers' productivity by forcing 
them to review alarms, diverting their attention from issues that matter (cf. surveys on 
alarms handling \cite{J:Heckman-Williams:2011,Muske-Serebrenik:2016}). 
Reducing and classifying those alarms based on previous
iterations/similarity significantly enhances the experience of using \SA tools.
This is a \emph{classification} \ML task: from a set of flagged alarms, which
ones are \emph{actionable}, i.e. require a specific bugfix from a developer.

Heckman \& Williams \cite{Heckman-Williams:2009} postulated that characterizing
whether an alarm is actionable highly depends on each project and developers
involved. They reduced the unactionable alarm number by first gathering alarms
and their features, then by selecting the most relevant ones for training a
classifier for future projects. They used as features the usual metrics (LoC,
etc.) with code change history, and alarm types (null pointers, etc.) with alarm
severity delivered by \SA tools.
%
Hanam \emph{et al.} \cite{Hanam-Tan-Holmes-Lam:2014} proposed to classify alarms
as actionable or not by identifying recurrent patterns based on characteristics
related to code statements: invocation and creation sites, field access, binary
operations, catch statements as well as other various structural features like
method signatures and class names.
They showed that patterns effectively exist and help discover more errors than
classical \SA tools reports.
Kremeneck \emph{et al.} \cite{Kremenek-Ashcraft-Yang-Engler:2004} correlated
alarm reports with their code localization to classify alarms raised at later
stages of code integration.
Ruthruff \emph{et al.} \cite{Ruthruff-etAl:2008} identified the legitimate
alarms that are more likely to be acted on by developers, based on \ML features
similar to \cite{Heckman-Williams:2009,Hanam-Tan-Holmes-Lam:2014}.
They also managed to reduce the number of metrics necessary for performing the
classification, while preserving a correct classification ratio.

\subsubsection{Predicting Bugs from Previous Code Versions}
\label{sec:SA-PredictingBugs}

Instead of running \SA tools during the development phase, an interesting line of research
consists of predicting, at code submission into a repository, whether a code change likely contains bugs, based on the analysis of previously submitted changes in a project.
This presents several benefits: the change is still fresh in mind, and several 
actions may be taken (from code review, testing, to \FV techniques), targeting the recent change. This is a \emph{classification} \ML task:
predict whether a code change is likely to contain bugs, based on the analysis
of previous code version(s).

Kim and his colleagues
\cite{J:Kim-Whitehead-Zhang:2008,J:Shivaji-Whitehead-Akella-Kim:2013} as well as
Hata \emph{et al.} \cite{Hata-Mizuno-Kikuno:2008} introduced change
classification by analyzing the change history, based on log messages keywords
and correlations to bug fix requests. Their predictors rely on various features:
change metadata on the versioning system, complexity metrics and various
\SA information to locate code change and analyze their impact.
Kim and his colleagues also investigated in \cite{J:Kim-Whitehead-Zhang:2008} the
possibility of reducing the large amount of features extracted from change
history.

\subsubsection{Classifying Android Apps as Malware} 
\label{sec:SA-Malware}

With hundreds of new apps and countless updates, detecting malware in Android apps has 
become crucial to ensure end-users' security. Most contributions mixing \ML with \SA rely on \emph{misuse detection} that flags an app when permissions mismatch the actual app functionalities (cf. \cite{Nath-Mehtre:2014} for an overview of static and dynamic malware detection). 
%
This qualifies as a 
\emph{classification} \ML task: given an Android app, together with a set of
characteristic features that are extracted statically, determine whether the app
contains a malware. The listed contributions differ in features and training data sizes for 
training: we comment on features and refer to each paper for other details.

Aung \& Zaw \cite{J:Aung-Zaw:2013} used five characteristic permissions
(internet access, configuration files change, send/write \textsc{Sms} and phone
calls) extracted directly from the distribution files (\textsf{.apk}) of known
malware and goodware from classical Android Markets.
Sahs \& Khan \cite{Sahs-Khan:2012} used a combination of permissions,
categorized as built-in (like accessing Internet) and non-standard (like
accessing the camera or the localization), paired with Control Flow Graphs to
analyze the app's code.
Yerima and his colleagues
\cite{J:Yerima-Sezer-Muttik:2014,Yerima-Sezer-Muttik:2015} (cf.
\cite{Yerima-Sezer-Muttik:2014} for details) 
extracted a set of complementary features,
characterized by specific keywords: a total of 125 permissions extracted from
the manifest; features related to Linux commands hidden in compiled or library
code, used for escalating privileges or launching scripts and malicious
binaries; and standard Android \textsc{Api} calls extracted from the app's
Dalvik code, to detect required interactions with the various devices (e.g.,
\textsc{Sim}, location or network accesses, device or user ids, or method
invocation and class loading, boot process, etc.) or to enrich apps with various
functionalities (e.g. contact, \textsc{Sms} or \textsc{Url} lists already
accessed).

\subsubsection{Learning SA}
\label{sec:SA-LearningSA}

Tackling the \SA problem itself, directly from the source program is a 
\emph{classification} \ML task: providing the \textsc{Ast}, does the (fixed, 
predefined) property hold or not. 

Several authors attempted this 
\cite{Chapelly-Cifuentes-Krishnan-Gevay:2017,Mou-Li-etAl:2016,Yamaguchi-Lottmann-Rieck:2012} 
for various analyses, but all noticed that, for the approach to scale, sufficient 
training data for each property needs to be available (positive as well as negative training, i.e. 
verifying \emph{and} falsifying the property at hand).

\section{Discussion}
\label{sec:Discussion}
\enlargethispage{12pt}

\begin{table*}%
	\begin{scriptsize}
\begin{center}
	\begin{tabular}{|c|l|l|l|l|}
		\cline{2-5}
		\nocell{1} & \textbf{Themes} & \textbf{Contributions} & \textbf{ML Task} &
		\textbf{\ML Model Types gathered from Contributions} \\
		\cline{2-5}
		\cline{1-5}
		\multirow{8}{*}{\rotatebox[origin=c]{90}{\textbf{\SAT-\textsc{Smt}}}} &
			Predicting Runtime               & \cite{Samulowitz-Memisevic:2007}(+) \cite{Horvitz-Ruan-etAl:2001}(-)  & Classification (or Regression) & Logistic Regression ; Decision Trees \\
			\cdashline{2-5}[.4pt/1pt]
			& \multirow{2}{*}{Restarting Computations} & 
				    \cite{Haim-Walsh:2009}(o)
				    \cite{Kautz-Horvitz-Ruan-etAl:2002,Ruan-Horvitz-Kautz:2002}(+) & Classification & Logistic Regression ; Decision Trees\\
				& & \cite{Nejati-Liang-Ganesh-Gebotys-Czarnecki:2017}(+) \cite{Gagliolo-Schmidhuber:2007}(o)   & Reinforcement  & Multi-Armed Bandit \\
			\cdashline{2-5}[.4pt/1pt]
			& Selecting Branching Variable       & \cite{Liang-Ganesh-Poupart-Czarnecki:2016a,Liang-Ganesh-Poupart-Czarnecki:2016b}(+) \cite{Liang-etAl:2017}(+) \cite{Frohlich-Biere-Wintersteiger-Hamadi:2015} \cite{J:Lagoudakis-Littman:2001}(+) & Reinforcement & Multi-Armed Bandit ; Temporal Difference\\
			\cdashline{2-5}[.4pt/1pt]
			& Determining Best-Solving Algorithm & \cite{J:Xu-Hutter-Hoos-LeytonBrown:2008}(+) \cite{J:Lindauer-Hoos-Hutter-Schaub:2015}(+)    & Classification  & Logistic Regression \\
			\cdashline{2-5}[.4pt/1pt]
			& Configuring Solvers' Parameters    & \cite{Hutter-Hamadi-Hoos-LeytonBrown:2006}(+) \cite{J:Lindauer-Hoos-Hutter-Schaub:2015}(+)  & Regression          & Support Vector Machines ; Random Forest Regression \\
			\cdashline{2-5}[.4pt/1pt]
			& Learning Satisfiability            & \cite{Wu:2017}(+) \cite{Xu-Hoos-LeytonBrown:2012}(+) \cite{Devlin-OSullivan:2008}(+) \cite{Selsam-Lamm-etAl:2018} & Classification  & \hspace{-0.2cm}\begin{tabular}{l}
				Logistic Regression ; Decision Trees ; Random Forests\\ 
				k-Nearest Neighbors ; Na\"ive Bayes ; Neural Network
				\end{tabular}\\
		\hline
		\multirow{5}{*}{\rotatebox[origin=c]{90}{\textbf{\TP}}} &
			Selecting Premises &
				\cite{AlemiCISU:16}(+) \cite{Loos:2017}(+)  \cite{Fuchs1998}(+) \cite{Alama:2011}(+) \cite{Kaliszyk2014}(+)
				\cite{Blanchette:2016}(+) \cite{KaliszykUV:14}(+) & Classification & Kernel-based models ; Na\"ive Bayes ; k-Nearest Neighbors\\
			\cdashline{2-5}[.4pt/1pt]
			& Configuring Proof Engine's Parameters & \cite{KuhlweinUrban:2015}(+) &
			Regression & Kernel-based models\\
			\cdashline{2-5}[.4pt/1pt]
			& Selecting Predefined Proof Engine Parameters    & \cite{Bridge2010}(+)
			\cite{Bridge2014}(+)  & Classification & Gaussian Process Classifiers;
			Kernel-based models\\
			\cdashline{2-5}[.4pt/1pt]
			& Guiding Interactive Proofs         & \cite{Urban:2006}($\circ$)
			\cite{Mercer:2006}($\circ$)  \cite{Komendantskaya:2012}($\circ$) &
			Classification / Clustering  & Na\"ive Bayes ; Variable Length Markov Models / k-Means\\
			\cdashline{2-5}[.4pt/1pt]
			& Learning Theorem-Proving           & \cite{Rocktaschel:2017}(-)& Regression & Neural Networks\\
		\hline
		\multirow{3}{*}{\rotatebox[origin=c]{90}{\textbf{\MC}}} &
			Finding Counterexamples                 & \cite{Araragi-MoCho:2006}(+) \cite{Behjati-Sirjani-Ahmadabadi:2009}(+)     & Reinforcement     & Q-Learning \\
			\cdashline{2-5}[.4pt/1pt]
			& Refining Abstractions                 & \cite{Clarke-Gupta-Kukula-Strichman:2002,J:Clarke-Gupta-Strichman:2004}(+) & Classification    & Decision Trees \\ 
			\cdashline{2-5}[.4pt/1pt]
			& Extracting Most Common Error Patterns & \cite{J:Pira-Rafe-Nikanjam:2016}(o)                                        & Frequent Item Set & A-Priori \\
		\hline
		\multirow{6}{*}{\rotatebox[origin=c]{90}{\textbf{\SA}}} & 
			\multirow{2}{*}{Identifying Actionable Alerts}                 &
			\multirow{2}{*}{\cite{Heckman-Williams:2009}($\circ$) \cite{Hanam-Tan-Holmes-Lam:2014}(+)
			\cite{Kremenek-Ashcraft-Yang-Engler:2004}(+) \cite{Ruthruff-etAl:2008}(+)} & 
			\multirow{2}{*}{Classification} & 
			Decision Trees ; Bayesian models ; \\
			& & & & Logistic Regression ; Rule-Based\\
			\cdashline{2-5}[.4pt/1pt]
			& Predicting Bugs from Previous Code Versions &
			\cite{J:Kim-Whitehead-Zhang:2008}($\circ$)
			\cite{J:Shivaji-Whitehead-Akella-Kim:2013}(+)
			\cite{Hata-Mizuno-Kikuno:2008}(+) & Classification & Support Vector Machines ; Bayesian models \\
			\cdashline{2-5}[.4pt/1pt]
			& \multirow{2}{*}{Classifying Android Apps as Malware}         &
			\multirow{2}{*}{\cite{J:Aung-Zaw:2013}($\circ$) \cite{Sahs-Khan:2012}($\circ$) \cite{J:Yerima-Sezer-Muttik:2014}(+) \cite{Yerima-Sezer-Muttik:2015}(-)} &
			\multirow{2}{*}{Classification/Clustering} & Decision Trees ; Random Forests ; \\
			& & & & Support Vector Machines ; Bayesian models / k-Means \\
			\cdashline{2-5}[.4pt/1pt] 
			& Learning \SA                                &
			\cite{Chapelly-Cifuentes-Krishnan-Gevay:2017}(-)
			\cite{Mou-Li-etAl:2016}(+)
			\cite{Yamaguchi-Lottmann-Rieck:2012}($\circ$)&
			Classification & Neural Networks
			\\
		\hline
	\end{tabular}
\end{center}
	\center{Legend: (+) improves the state of the art; (-) comparable to or worse
	than state of the art; ($\circ$) no information on how the approach
	relates to the state of the art}
	\end{scriptsize}
	\caption{Contributions and \ML tasks related to each theme within each \FV
	approach.}
	\label{tab:summary}
\end{table*}

Is there a love story between \FV and \ML? In this paper, we surveyed one
direction of this relationship: how \ML contributes to enhance \FV activities.
Without being exhaustive, we have shown throughout Section
\ref{sec:Contributions} that \ML enhances all the spectrum of \FV approaches
(Static Analysis, Model-Checking, Theorem-Proving, but also
\SAT/\textsc{Smt}-Solving) at different levels, using different techniques and
for different purposes. 

How can \FV and \ML experts collaborate to leverage \ML's current
practices in order to enhance and improve current \FV tools? We follow the
classical \ML pipeline:
(i) \emph{identifying the \FV problem} at hand; (ii) \emph{identifying the
corresponding \ML task}; (iii) \emph{providing \ML features} to characterize
the learning instances; (iv) \emph{figuring out which \ML model (type)} would
 perform adequately.

The remainder of this section discusses each of these points, gives general
perspectives on what seems promising for the future, and provide answers to the
research questions formulated in Section~\ref{sec:Protocol-RQ}.

\subsubsection*{FV Problems (\textbf{RQ1:} How is \ML used inside \FV tools?)}
\label{sec:Discussion-Problems}

\FV experts first identify what they expect to improve, compute, or which
kind of pattern they seek in their data. This is one of the topics covered by
this paper: each of the themes indicates precisely to which extent \ML is used
in the overall \FV approach: some contributions/tools invoke \ML at various
steps, or even handle the approach altogether. We noticed several patterns
according to \ML categories.
\emph{Supervised techniques} are often associated with two kinds of usage
observed in \SAT and \TP: \emph{external} guidance, which stands for \ML models that
choose an appropriate heuristic, strategy or algorithm (e.g., portfolio solving
in \SAT and proof engine parameter selection in \TP); and \emph{internal}
guidance, which represents situations where \ML models play the role of heuristics/strategies,
by selecting the next following step in a more global algorithm (e.g., fact
selection in \TP or restart selection in \SAT). Other uses do not fit these categories. 
For instance, the \emph{interpretative gap filling}
in \SA is performed through \ML models by
finding links or recurrent patterns in the collected information (e.g. the link between permissions and the malware/goodware classes).

\emph{Unsupervised techniques} do not prescribe an ``ideal'' answer, but rather
try to identify general patterns. The resulting tasks
(mostly clustering) seem more adequate for \TP, the only \FV
approach favoring interactivity. However, it is not excluded that unsupervised
techniques may bring new insights into other \FV approaches, even those that
already are fully automated.

\subsubsection*{ML Tasks (\textbf{RQ3:} What \ML task(s) is (are) used for
which purpose in \FV?)}
\label{sec:Discussion-Tasks}

Once the \FV problem is identified, \FV specialists associate an \ML task
to guide \ML experts towards the right set of techniques and models. When \ML
contributes to a small portion of the \FV process we have often observed
\emph{classification} tasks, i.e. selecting a candidate artifact among several
available ones. In \TP and \SAT, selecting heuristics according to some criteria
(potential for proof completion; solving runtime) among those already programmed
by experts, relieves \FV users from the burden of having to maintain explicit
knowledge of those heuristics. Such approaches have improved tools
significantly, as witnessed by Tool Contests in \SAT and \TP.
In \MC and \SAT, \emph{reinforcement learning} is used as a way to ``guide'' the tool
towards a counterexample and the most promising branching variable.
The final \ML task may differ according to the experts' viewpoint, but is
ultimately guided by \FV experts' needs: for example, a \emph{regression} task such
as predicting \SAT solvers' parameters values, or runtimes in \TP, may very well 
be turned into a \emph{classification} task by imposing runtime thresholds.

\subsubsection*{ML Features (\textbf{RQ5:} How are \ML features extracted/selected to
guide \ML tasks?)}
\label{sec:Discussion-Features}

The choice of features is critical for the learning process. We noticed two main
categories of such choices in \FV: features are either based on the \emph{raw
input} used for the \FV approach (e.g., \textsc{Ast} and manifests in \SA;
\textsc{Cnf}s in \SAT and \TP); or based on \emph{measures} computed on those
raw inputs (e.g. Call Graphs for \SA, ratios and clause numbers for \SAT). Those
features were identified experimentally while tuning heuristics and/or trying to
improve existing algorithms, and often predate the introduction of \ML.
Identifying the appropriate features is the task of \FV experts, but using \ML
may help enhancing them or filtering out superfluous ones. 
Table \ref{tab:features} lists the main contributions in each \FV approaches.

\begin{table}%
    \renewcommand{\arraystretch}{1.8}
	\begin{footnotesize}
\begin{center}
	\begin{tabular}{cc}
		& \textbf{Reference Papers} \\
		\hline
		\textbf{\SAT-\textsc{Smt}} & 
			\cite{J:Xu-Hutter-Hoos-LeytonBrown:2008,Xu-Hoos-LeytonBrown:2007}
			\cite{J:Ansotegui-Bonet-GiraldezCru-Levy:2016}
		\\
		\textbf{\TP} & 
			\cite{meng:2009} \cite{Kaliszyk2014,Kaliszyk:2015} 
			\cite{Kuhlwein2013} \cite{KuhlweinUrban:2015}
		\\
		\textbf{\MC} & 
			\cite{J:Clarke-Gupta-Strichman:2004}
		\\
		\textbf{\SA} & 
			\cite{Heckman-Williams:2009,Hanam-Tan-Holmes-Lam:2014}
			\cite{J:Shivaji-Whitehead-Akella-Kim:2013} \cite{Hata-Mizuno-Kikuno:2008} 			
			\cite{J:Aung-Zaw:2013} \cite{Sahs-Khan:2012} \cite{Yerima-Sezer-Muttik:2015}
		\\
		\hline
	\end{tabular}
\end{center}
	\end{footnotesize}
	\caption{Main papers defining \ML features.}
	\label{tab:features}
\end{table}

\subsubsection*{ML Models (\textbf{RQ4:} Which \ML model type are used to 
perform the \ML tasks?)}
\label{sec:Discussion-Models}

The selection of an \ML model to perform a specific \ML task is the final step.
This is extremely delicate, and essentially a problem for \ML experts. We
noticed in the surveyed contributions that the use of specific models by \FV experts is
not always clearly motivated. In fact, the literature suggests that the \ML
model is often selected among the set of those available in the \ML tool(s) the
authors are familiar with (e.g. the Weka Workbench
\cite{B:Frank-Hall-Witten:2016}). 
This is not incompatible with common practice: \ML experts often use a trial-and-error
approach to determine which model performs best for a given task.
Having bad performance with a specific \ML model does not always imply that the model is
not suited for the \ML task, but rather it is not optimally parameterized. 
It is sometimes impossible to figure out in advance which model will work best 
(as a consequence of Wolpert and McReady's No Free Lunch Theorem \cite{J:Wolpert-McReady:1997}). 
However, we believe that identifying precisely the answers to the previous steps 
should provide \ML practitioners with sufficient information such that they can 
exercise their expertise. 

\subsubsection*{\textbf{RQ2:} Is using \ML inside \FV tools beneficial?}
\label{sec:Discussion-Improvement}

Table~\ref{tab:summary} presents a summary of our findings: for each theme inside 
each \FV approach, we list the contributions we reviewed and indicate whether the
results of each contribution has improved the existing state-of-the-art, and points
to the models commonly used by all contributions in a theme. 
From a statistical viewpoint, contributions that claim to have brought 
improvements largely outnumber the ones with similar or lower quality than 
state-of-the art. 
\subsubsection*{Towards end-to-end \FV}
\label{sec:Discussion-E2E}

All in all, we observed a pyramidal use of \ML models. On one end of the
spectrum, \ML models are used in a very narrow fashion inside \FV tools for
solving very specific problems inside tools. For instance, in \SAT or \TP, the
structure of the current resolution algorithms can be preserved while delegating
onto \ML models the optimization of specific choices that are
traditionally handled by heuristics (like restarts or fact selections).
On the other end of the spectrum, we noticed several attempts to handle an \FV
approach globally: for instance, predicting satisfiability of a formula
\cite{Wu:2017,Xu-Hoos-LeytonBrown:2012}, building a proof
\cite{Rocktaschel:2017} or learning static analysis directly
\cite{Chapelly-Cifuentes-Krishnan-Gevay:2017,Mou-Li-etAl:2016,Yamaguchi-Lottmann-Rieck:2012}.
Between these two ends, a range exists determined by how much \FV expert
knowledge is considered while solving the \FV problem. \ML typically aims at finding objective generalizations;
however, if injecting domain knowledge (e.g. on how current \FV algorithms
are designed) significantly improves performances, it becomes meaningful to
integrate it to relieve the \ML algorithms from struggling to learn specific
aspects while allowing them to focus on more global aspects of the problem. Our survey points to that fact that the use of \ML for specific \FV tasks 
is over-represented, at the expense of more recent holistic strategies.

Such holistic strategies have already radically changed fields such as image
and natural language processing, especially after the introduction of Deep
Learning \cite{B:Goodfellow-Bengio-Courville:2016}.
It became clear from our literature study that \FV tools that introduced \ML in
their workings started to deliver impressive performances in international
contests (e.g. \cite{Loos:2017,Bridge2014} in \TP;
\cite{J:Xu-Hutter-Hoos-LeytonBrown:2008,J:Lindauer-Hoos-Hutter-Schaub:2015} in \SAT).
However, using more powerful \ML models directly hampers the interpretability of
their results~\cite{Bibal:2016}. This is problematic for \FV, since its
techniques are often used to ensure the correctness of safety-critical software
for which human-understandable justifications need to be provided.
We strongly call for a more systematic review of the domain in order to
precisely identify future directions in this research domain. We believe
\emph{machine learned} \FV is potentially achievable when sufficient amounts of data
will be collected, just like image analysis for medical diagnostic
\cite{J:WeiTing-Cheung-Lim:2017}, board game playing 
\cite{Silver:2016,Silver:2017} or even self-driving cars
\cite{Bojarski-etAl:2016,Bojarski-etAl:2017}. While a decade ago progress in
such domains seemed extremely difficult, it has now become reachable for
(state-of-the-art) \ML.
Rather than following designs and abstractions created by humans, \ML may indeed
find fresh new ways of handling \FV problems, opening the potential to
entirely reshaping the \FV domain.

\section{Threats to Validity}
\label{sec:validity}

Although we designed our search protocol in a very inclusive way, as witnessed by
the number of papers finally discarded manually, we may have missed some relevant
contributions. First, we relied exclusively on repository search, whereas crossing
searches with top-venues both in \FV and \ML may have brought new interesting 
contributions. Second, we stopped searching online repository after 10 result pages 
Third, we performed the search in September 2017, before many important \FV, as well as \ML, 
venues take place. To mitigate these points, we have conducted a backward snowballing
on the main papers in each approach (typically, the most cited ones) in January 2018,
and looked at the program of some of the relevant top venues. In future
revisions of our survey, we will integrate forward snowballing, as recommended by 
Wohlin \cite{Wohlin:2014}, by looking \emph{a posteriori} at top-venues in recent months. 

Furthermore, our survey is likely to have missed gaps in the literature, 
meaning that additional themes and/or better relations between themes,
tasks and \ML models may surface in future work. In fact, as research progresses
in this area and \ML becomes more widely adopted, we expect to find new themes that
are for the moment not explored by the community: this survey may well be
seen as a current snapshot of the available contributions in the domain, rather
than a definitive survey that closes the matter.

In Table \ref{tab:summary}, the information regarding state-of-the-art
improvements (noted as +/o/-) has been collected from each contribution relying
solely on the article's text. We have taken into account the comparisons with
other tools operated by the authors, or analyzed explicit statements from them
on how their method/tool compare to others. We have been particularly attentive
to available data on relevance (precision and recall) and performance (speed).
These comparisons found in the literature form a heterogeneous set: some authors
compare their work with solutions where no \ML is used, whereas others provide comparisons with
\ML-based tools; datasets which are used as the basis for learning are often
small and have not been made available online, meaning reproducibility of the
presented results is, in general, not possible. In some articles, no comparison
with the state of the art is provided by the authors:
in some instances, this means that the \ML technique addresses a problem that
was previously manually handled, or not handled at all; in others, this simply means
no comparison is provided by the authors.

Finally, the \ML model types
presented in Table~\ref{tab:summary} have been gathered for informational
purposes and not in an exhaustive way. In this sense, the association between
themes and \ML models does not imply an exclusive relation of appropriateness
between them.

\section{Conclusion}
\label{sec:Conclusion}

In this paper, we have explored how \ML contributes to enhance \FV tools efficiency.
We covered four classical approaches, namely \SAT, Theorem-Proving (\TP), 
Model-Checking (\MC), and Static Analysis (\SA), for which we catalog a list of 
themes, i.e. precise points where an \FV problem is translated into an \ML task
to be handled by an \ML model type. 
Although preliminary, our survey shows not only that \ML 
may keep on contributing
to the \FV field in both short and medium terms. It also shows that integrating \ML
methods inside \FV tools is largely beneficial, as demonstrated in \SAT and \TP 
tools that regularly achieve new scales of efficiency. However it is still essential to 
tackle challenges that were until recently thought as unreachable (e.g. 
attacking realistic cryptography protocols like \textsc{Rsa}). 

By studying the intricate relation between \FV and \ML over a large spectrum of
approaches, we were able to frame the way \FV and \ML experts collaborate in 
a classical \ML pipeline: identifying the \FV problem (corresponding to our 
\emph{themes}); determining the corresponding \ML task; providing \ML features; 
and figuring out which \ML model type is the most adequate for the task. We also 
captured general trends, the most challenging being \emph{learning} \FV approaches 
on their own, as witnessed by many attempts in e.g. \SAT, \TP and \SA.




The reverse direction of the love story has been left untouched, despite a recent 
growing interest: how can \FV may help \ML. 
Verifying \ML tasks results has nowadays become a stepping stone in the adoption
of thrilling new technologies: for example, correctly identifying road signs 
directly influences the behavior of self-driven cars, which in turns guarantees
the safety of passengers. One of the main reasons that make such verification hard
is that the implicit models (e.g. neural networks, one of the currently most 
promising learning technologies) are difficult to grasp and understand for
humans. Properly stating what kind of properties one expects from
such implicit models is even more difficult. Building appropriate abstractions
of such models, that often integrate probabilistic and/or continuous
computations is a key challenge. Therefore, specifying what models to accept
 appears to be difficult. 



\balance
\bibliographystyle{abbrv} 
\bibliography{./arXiv2018}

\end{document}